\documentclass{emulateapj}

\usepackage{graphics}
\usepackage{epsfig}
\usepackage{natbib}
\shorttitle{The UV Spectrum of GJ 876}
\shortauthors{France et al.}
\begin{document}

\title{Time-resolved Ultraviolet Spectroscopy of the M-dwarf GJ 876 Exoplanetary System\altaffilmark{*}}


\author{
Kevin France\altaffilmark{1},
Jeffrey L. Linsky\altaffilmark{2}, 
Feng Tian\altaffilmark{3,4},
Cynthia S. Froning\altaffilmark{1}, 
Aki Roberge\altaffilmark{5}
}

\altaffiltext{*}{Based on observations made with the NASA/ESA $Hubble$~$Space$~$Telescope$, obtained from the data archive at the Space Telescope Science Institute. STScI is operated by the Association of Universities for Research in Astronomy, Inc. under NASA contract NAS 5-26555.}

\altaffiltext{1}{Center for Astrophysics and Space Astronomy, University of Colorado, 389 UCB, Boulder, CO 80309; kevin.france@colorado.edu}
\altaffiltext{2}{JILA, University of Colorado and NIST, 440 UCB, Boulder, CO 80309}
\altaffiltext{3}{Current Address: Center for Earth System Sciences, Tsinghua University, Beijing, China 100084}
\altaffiltext{4}{Laboratory for Atmospheric and Space Physics, University of Colorado, Boulder, CO 80309 }
\altaffiltext{5}{Exoplanets and Stellar Astrophysics Laboratory, 
	NASA Goddard Space Flight Center, Greenbelt, MD 20771}	


\begin{abstract}

Extrasolar planets orbiting M-stars may represent our best chance to discover habitable worlds in the coming decade.  The ultraviolet spectrum incident upon both Earth-like and Jovian planets is critically important for proper modeling of their atmospheric heating and chemistry.  
In order to provide more realistic inputs for atmospheric models of planets orbiting low-mass stars, we present new near- and far-ultraviolet (NUV and FUV) spectroscopy of the M-dwarf exoplanet host GJ 876 (M4V).   Using the COS and STIS spectrographs aboard the {\it Hubble Space Telescope}, we have measured the 1150~--~3140~\AA\ spectrum of GJ 876.  We have reconstructed the stellar \ion{H}{1} Ly$\alpha$ emission line profile, and find that the integrated Ly$\alpha$ flux is roughly equal to the rest of the integrated flux (1150~--~1210~\AA\ + 1220~--~3140~\AA) in the entire ultraviolet bandpass ($F$(Ly$\alpha$)/$F$(FUV+NUV)~$\approx$~0.7).  This ratio is $\sim$~2500$\times$ greater than the solar value.  We describe the ultraviolet line spectrum and report surprisingly strong fluorescent emission from hot H$_{2}$ ($T$(H$_{2}$)~$>$~2000 K).   We show the light-curve of a chromospheric + transition region flare observed in several far-UV emission lines, with flare/quiescent flux ratios $\geq$~10.   The strong FUV radiation field of an M-star (and specifically Ly$\alpha$) is important for determining the abundance of O$_{2}$~--~and the formation of biomarkers~--~in the lower atmospheres of Earth-like planets in the habitable zones of low-mass stars.     

\end{abstract}

\clearpage

\section{Introduction}

The advent of large radial velocity surveys and the $Kepler$ mission~\citep{udry07,borucki11} have revealed planets with masses several times that of Earth in orbits that may permit the existence of liquid surface water~\citep{kasting93}.  Perhaps the most favorable objects for follow-up studies in the next decade are planets in the habitable zone (HZ) around low-mass stars, dwarfs of spectral type M0V and later (also referred to as M-dwarfs or dM stars).  These planets have received considerable theoretical attention, and a consensus is emerging that neither synchronous rotation nor close proximity to the host star rules out the possibility of biological development on these worlds~\citep{joshi03,segura05,wordsworth10}. 

The heating and chemistry of an Earth-like atmosphere depends strongly on the characteristics of the host star, both spectrally and temporally.  The  photospheric ultraviolet (UV) continuum of M-stars is very small relative to solar-type stars due to their lower effective temperature.  Bright chromospheric and transition region emission lines dominate the UV spectrum of M-dwarfs, containing a much larger fraction of the stellar luminosity than for solar-type stars.  These differences influence the atmospheric temperature profile and the production of potential biomarkers (e.g. ozone, O$_{3}$; Segura et al. 2003; Segura et al. 2005).\nocite{segura03,segura05}  H$_{2}$O, CH$_{4}$, and CO$_{2}$ are highly sensitive to far-UV radiation, with photo-absorption cross-sections that overlap with \ion{H}{1} Ly$\alpha$, the strongest emission feature in the UV spectrum of an M-star.  M-stars are known to display frequent and intense flare activity~\citep{hawley96,west04,welsh07}, and the combination of UV photon and proton bombardment from a large flare may produce  $\gtrsim$~90\% O$_{3}$ depletions in an Earth-like atmosphere~\citep{segura10}.  Despite the importance of understanding these host M-stars, their actual spectral and temporal behavior is not well known outside of a few active flare stars~\citep{linsky94,pagano00,hawley03,osten05}.  

GJ~876 is an M4V star ($M_{*}$~=~0.33~$M_{\odot}$) hosting a multi-planet system~\citep{delfosse98,marcy98,marcy01,rivera05,rivera10,correia10}.  
The four planets have masses ranging from super-Jovian to super-Earth ($M_{b}$ = 2.64 $M_{Jup}$, $a_{b}$ = 0.21 AU; $M_{c}$ = 0.83~--~0.86 $M_{Jup}$, $a_{c}$ = 0.13 AU; $M_{d}$ $\approx$ 6 $M_{\oplus}$, $a_{d}$ = 0.02 AU; $M_{e}$ = 12 $M_{\oplus}$, $a_{e}$ = 0.33 AU), with GJ 876b orbiting near the center of the HZ.   Conditions favorable for the sustained presence of liquid surface water might therefore exist on any satellites of GJ 876b. There is no present evidence for a bright remnant debris disk around GJ 876~\citep{shankland08}, and taken together with the slow stellar rotation (40~$\leq$~$P_{*}$~$\lesssim$~97 days; Rivera et al. 2005, 2010), GJ 876 is estimated to have an age between 0.1~--~5 Gyr~\citep{correia10}.   On the basis of its H$\alpha$ absorption spectrum, GJ 876 is considered to be an inactive M-dwarf.  However, GJ 876 has a measureable level of coronal emission in soft X-rays ($L_{X}$~$\sim$~3~$\times$~10$^{26}$ erg s$^{-1}$ in the 0.2~--~2.0 keV band; Poppenhaeger et al. 2010).~\nocite{poppenhaeger10}  \citet{walkowicz08} have further challenged the assumed link between optical activity indicators and UV flux, using low-resolution $HST$-ACS prism observations to demonstrate that GJ 876 has levels of chromospheric near-UV (NUV) flux comparable to the well-studied active M-dwarf AD Leo.   

Other than the low-resolution spectra presented by~\citet{walkowicz08}, there are very few NUV observations of GJ 876, and none that clearly detect emission from the star at far-UV wavelengths (FUV; 1150~$<$~$\lambda$~$<$~2000~\AA).   As current M-dwarf models do not properly account for chromospheric and transition region emission, the lack of observational basis means that a crucial input to atmospheric models of planets orbiting low-mass stars is missing.   The situation is sufficiently unclear that atmospheric modelers have taken widely varying approaches to estimating the UV radiation field incident upon the HZ of M-dwarfs, from assuming the UV spectrum of a young flare star (AD Leo; Wordsworth et al. 2010) to assuming an effective UV flux of zero~\citep{kaltenegger11}.  As suggested by the latter authors, constraints from direct observation are required.~\nocite{wordsworth10}   In order to address this situation, we have begun to acquire MUSCLES (Measurements of the Ultraviolet Spectral Characteristics of Low-mass Exoplanetary Systems).   In this Letter, we present our first results from MUSCLES; NUV and FUV spectra of GJ~876 obtained with the {\it Hubble Space Telescope}-Cosmic Origins Spectrograph (COS) and Space Telescope Imaging Spectrograph (STIS).  

\begin{figure}
\begin{center}
\epsfig{figure=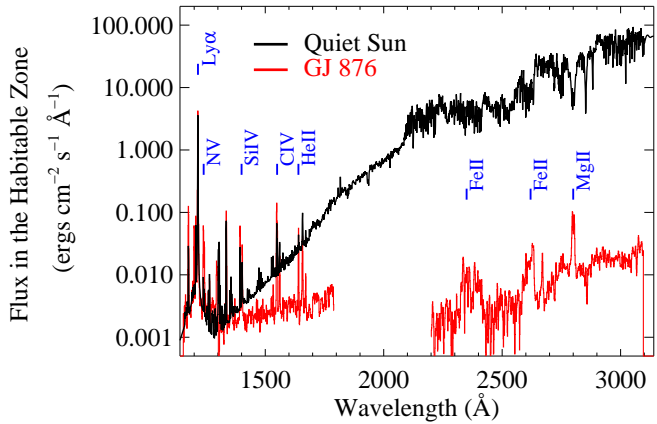,width=2.5in,angle=90}
\vspace{-0.2in}
\caption{
\label{cosovly} A comparison of the solar 1150~--~3140~\AA\ flux incident on the habitable zone of the Solar System (solar flux at 1 AU, $black$) and the observed flux from GJ 876 in the HZ (at 0.21~AU, $red$).  The 1800~--~2200~\AA\ region for GJ 876 has been excluded due to low S/N.  The 1150~--~1450~\AA\ (excluding Ly$\alpha$) spectrum is averaged over the flare and quiescent observations. 
 }
\end{center}
\end{figure}

\section{$HST$ Observations}

GJ~876 was observed with COS and STIS on 05 January 2012 and 12 November 2011, respectively ($HST$ Program ID 12464).  Multiple central wavelengths and focal-plane offset (FP-POSs) settings with the COS G130M and G160M modes 
were used to create a continuous FUV spectrum from 1143~--~1798~\AA.  These modes provide a point-source resolution of $\Delta$$v$~$\approx$~17 km s$^{-1}$ with 7 pixels per resolution element~\citep{osterman11}.  
The total exposure times in the G130M and G160M modes are 2017s and 2779s, respectively.  

Because COS is a slitless spectrograph, observations at \ion{H}{1} Ly$\alpha$ are heavily contaminated by geocoronal emission.  We therefore observed the stellar Ly$\alpha$ profile with STIS, using the G140M/cenwave 1222 mode ($\Delta$$v$~$\approx$~30 km s$^{-1}$) mode through the 52\arcsec~$\times$~0.1\arcsec\ slit for a total of 1138s.  Using the G230L mode on STIS, we also acquired the 1800~--~3140~\AA\ spectrum of GJ~876 at $\Delta$$v$~$\lesssim$~400 km s$^{-1}$.  

\begin{figure}
\begin{center}
\epsfig{figure=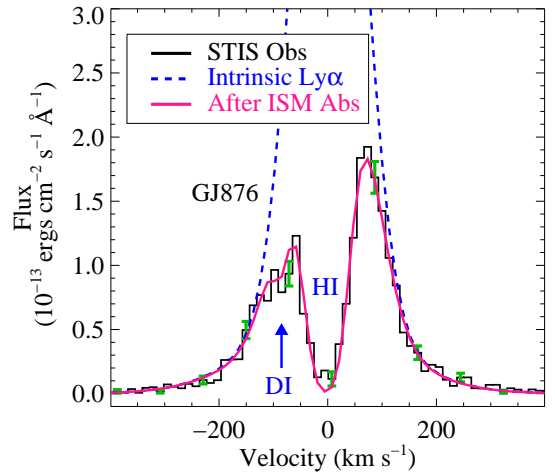,width=2.75in,angle=90}
\vspace{-0.2in}
\caption{
\label{cosovly} Velocity profile of the stellar Ly$\alpha$ emission line.  The STIS G140M observation of GJ876 
is displayed as the black histogram, with representative error bars plotted in green.  The Ly$\alpha$ line profile was reconstructed by assuming a 2-component Gaussian emission line ({\it blue dashed line}) and correcting for interstellar neutral hydrogen and deuterium ($N$(HI) = 1.1~$\times$~10$^{18}$ cm$^{-2}$, $b$~=~8.0 km s$^{-1}$, D/H~=~1.5~$\times$~10$^{-5}$; the final model is shown in $magenta$).  
}
\end{center}
\end{figure}

\section{Spectrally and Temporally Resolved Ultraviolet Spectra of GJ 876} 

\subsection{The UV Spectrum of GJ~876}

In Figure 1, we display the full 1150~--~3140~\AA\ UV spectrum of GJ 876 ($red$).  We observe emission from atomic and ionic species tracing a range of 
formation temperatures, including Ly$\alpha$, \ion{C}{2}, \ion{Al}{2} ($T$~$\sim$~1~--~3~$\times$~10$^{4}$ K); \ion{C}{3}, \ion{Si}{3}, \ion{Si}{4}, \ion{He}{2} ($T$~$\sim$~6~--~8~$\times$~10$^{4}$ K); \ion{C}{4}, \ion{N}{5}, and \ion{O}{5} ($T$~$\sim$~1.0~--~1.6~$\times$~10$^{5}$ K).  These emission lines clearly indicate that there is an active chromosphere and transition region on this star.  As suggested by~\citet{walkowicz08}, GJ 876 is $not$ an inactive star when observed in the UV.   GJ 876 may be in the class of ``weakly active'' stars whose optical spectra show H$\alpha$ in absorption while displaying \ion{Ca}{2} H~\&~K emission lines~\citep{rauscher06,walkowicz09}.  
The STIS G230L observations show emission lines from \ion{Mg}{2} and \ion{Fe}{2} as well as a weak NUV continuum at $\lambda$~$>$~2200~\AA.  
In Figure 1, we plot the flux observed at the orbit of GJ 876b,  roughly centered in the HZ of GJ 876 ($F_{HZ}$ = $F_{obs}$ $\times$~($d$/$a_{b}$)$^{2}$; $a_{b}$~=~0.21 AU).  The HZ spectrum of GJ 876 is compared with the Solar spectrum (Woods et al. 2009). \nocite{woods09} The GJ 876 data have been convolved with a 2~\AA\ FWHM Gaussian kernel for comparison with the Solar data.   Table 1 presents the band- and line-integrated HZ fluxes for several key regions/species.  

\subsection{Reconstruction of the \ion{H}{1} Ly$\alpha$ Emission Line Profile}

In order to produce an accurate estimate of the Ly$\alpha$ flux incident on the planets in the GJ~876 system, the Ly$\alpha$ profile observed by STIS must be corrected for attenuation by interstellar hydrogen and deuterium (e.g., Ehrenreich et al. 2011).\nocite{ehrenreich11}  In order to do this, we approximate the intrinsic Ly$\alpha$ emission line as a two-component Gaussian (as suggested for transition region emission lines from late-type stars by Wood et al. 1997) 
 and create a simultaneous least-squares fit to the stellar emission and interstellar absorption components.~\nocite{wood97}  The Ly$\alpha$ emission components are characterized by an amplitude, FWHM, and velocity centroid ($v^{narrow,broad}_{emis}$).  Interstellar absorption is parameterized by the column density of neutral hydrogen ($N$(HI)), Doppler $b$ parameter, D/H ratio, and the velocity of the interstellar absorbers ($v_{ISM}$).  We assumed a fixed D/H ratio (D/H = 1.5~$\times$~10$^{-5}$; Linsky et al. 1995; 2006).\nocite{linsky95,linsky06}   

A two-component fit with FWHMs~=~133 and 303 km s$^{-1}$, and an interstellar atomic hydrogen absorber characterized by log$_{10}$($N$(\ion{H}{1}))~=~18.06 and $b$~=~8.0 km s$^{-1}$, is able to produce an adequate fit ($\chi^{2}_{\nu}$~=~1.73) to the observations.   
The best model has $v^{narrow}_{emis}$ = +2.4 km s$^{-1}$, $v^{broad}_{emis}$  = $-$8.0 km s$^{-1}$, and $v_{ISM}$ = $-$3.0 km s$^{-1}$; the Ly$\alpha$ emission is somewhat blueshifted relative to the +8.7 km s$^{-1}$ radial velocity of the star.   The ISM velocity from our best-fit model is consistent with the expected Local Interstellar Cloud velocity towards GJ 876 ($v_{LIC}$~=~-3.56 km s$^{-1}$; Redfield \& Linsky 2008), supporting our Ly$\alpha$ reconstruction analysis.\nocite{redfield08} 
A single Gaussian Ly$\alpha$ emission line is unable to reproduce the observed profile ($\chi^{2}_{\nu}$~$>$~6). The reconstructed stellar Ly$\alpha$ profile is shown in Figure 2.  Integrated over the model profile, the total intrinsic Ly$\alpha$ flux is 4.38~$\times$~10$^{-13}$ erg cm$^{-2}$ s$^{-1}$.   The uncertainty on the integrated intrinsic flux is $\approx$~5~--~10~\%, while the uncertainty on any individual fit parameter is most likely $\approx$~20~\%.  We note that the correction for interstellar scattering does not take into account higher order opacity sources such as the stellar wind and heliosphere, which are not expected to significantly affect our results~\citep{wood05}.  

\begin{figure}
\begin{center}
\epsfig{figure=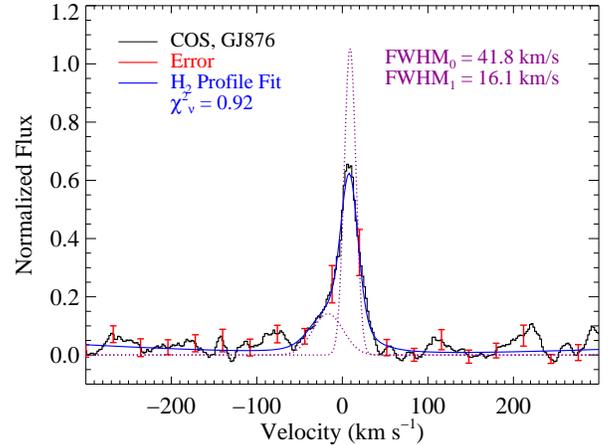,width=2.6in,angle=90}
\vspace{-0.2in}
\caption{
\label{cosovly} 
Co-added velocity profile of six fluorescent H$_{2}$ emission lines observed by COS.   
The narrow emission component is unresolved (FWHM$_{H2}$ = 16.1 km s$^{-1}$) and consistent with the stellar radial velocity of +8.7 km s$^{-1}$ ($v_{H2}$ = 9.0 km s$^{-1}$). 
The weaker, broad component (offset by -26 km s$^{-1}$) is only marginally detected.   
The dotted lines are the intrinsic Gaussian H$_{2}$ emission lines and the overplotted solid blue line is the H$_{2}$ spectral fit convolved with the COS LSF.   The narrow component is consistent with an origin in the stellar atmosphere, and 
the broad H$_{2}$ component, if real, may be attributed to bulk motion in the stellar atmosphere, or possibly, molecular gas outflowing from the giant planets in the GJ 876 system.  
 }
\end{center}
\end{figure}

\begin{deluxetable*}{c|ccc}
\tabletypesize{\normalsize}
\tablecaption{Habitable Zone UV Fluxes. \label{lya_lines}}
\tablewidth{0pt}
\tablehead{
\colhead{} & \colhead{$\Delta$$\lambda$\tablenotemark{a}}  &  \colhead{Solar\tablenotemark{b}} & 
\colhead{GJ876\tablenotemark{c}}  \\
\colhead{} & \colhead{ (\AA)}  &  \colhead{(ergs cm$^{-2}$ s$^{-1}$)} & 
\colhead{(ergs cm$^{-2}$ s$^{-1}$)} 
}
\startdata
$F$(Ly$\alpha$)\tablenotemark{d} & (1210~--~1220) & 5.6~$\times$~10$^{0}$  & 9.5~($\pm$~0.7)~$\times$~10$^{0}$ \\
$F$(\ion{N}{5})\tablenotemark{e} & (1235~--~1245) & 1.8~$\times$~10$^{-2}$  & 2.0~($\pm$~0.1)~$\times$~10$^{-1}$  \\
$F$(\ion{Si}{4})\tablenotemark{e} & (1390~--~1410) & 5.6~$\times$~10$^{-2}$  & 1.9~($\pm$~0.1)~$\times$~10$^{-1}$  \\
$F$(\ion{C}{4}) & (1545~--~1555) & 1.2~$\times$~10$^{-1}$  & 4.2~($\pm$~0.3)~$\times$~10$^{-1}$  \\
$F$(\ion{He}{2}) & (1635~--~1645) & 2.4~$\times$~10$^{-2}$  & 1.2~($\pm$~0.1)~$\times$~10$^{-1}$  \\

$F$(FUV) & (1150~--~1210) + & 1.8~$\times$~10$^{1}$  & 4.1~($\pm$~0.6)~$\times$~10$^{0}$  \\
		 & (1220~--~1790) &  &   \\
$F$(NUV) & (2200~--~3140) & 2.2~$\times$~10$^{4}$  & 9.8~($\pm$~0.6)~$\times$~10$^{0}$   \\
\tableline
\tableline
$F$(Ly$\alpha$)/$F$(FUV) &   & 0.3  & 2.3   \\
$F$(Ly$\alpha$)/$F$(FUV+NUV) &  & 2.6~$\times$~10$^{-4}$  & 0.7   \\
\enddata
\tablenotetext{a}{$\Delta$$\lambda$ is the bandpass over which the flux is calculated. } 
\tablenotetext{b}{Observed Solar flux at 1 AU. } 
 \tablenotetext{c}{Observed UV Flux scaled to the orbit of GJ 876b} 
\tablenotetext{d}{Intrinsic Ly$\alpha$ flux after correction for interstellar absorption.} 
\tablenotetext{e}{FUV flare contributes $\sim$~50\% and 25\% of the tabulated emission line fluxes of \ion{N}{5} and \ion{Si}{4}, respectively. } 
\end{deluxetable*}

\subsection{H$_{2}$ Fluorescence} 
A surprising result from the COS G160M observations is the detection of relatively strong molecular hydrogen fluorescence.  We clearly detect six H$_{2}$ emission lines pumped by Ly$\alpha$ through the (1~--~2) R(6) $\lambda$1215.73~\AA\  and (1~--~2) P(5) $\lambda$1216.07~\AA\ transitions.  The lines pumped through (1~--~2) R(6) are: [(1~--~7)R(6) $\lambda$1500.24~\AA\ and  (1~--~7)P(8) $\lambda$~1524.65~\AA].  The lines pumped through (1~--~2) P(5) are: [(1~--~6)R(3) $\lambda$1431.01~\AA, (1~--~6)P(5) $\lambda$~1446.12~\AA,(1~--~7)R(3) $\lambda$1489.57~\AA, and (1~--~7)P(5) $\lambda$~1504.76~\AA].  The Ly$\alpha$-pumping mechanism requires a significant population of molecules in the $v$~=~2 level of the ground electronic state, typically implying rotational temperatures of $T$(H$_{2}$)~$>$~2000K  (see Herczeg et al. 2004 and France et al. 2010a).\nocite{herczeg04,france10b}  The relative H$_{2}$/atomic line ratios are large compared with other M-stars (AD Leo, Proxima Cen, EV Lac, and AU Mic).  Since the S/N of the individual H$_{2}$ lines is relatively low, we coadded the six cleanly detected lines (Figure 3).
We present a discussion of the origin of the H$_{2}$ emission in \S4.2.

\subsection{FUV Flare}

During the COS G130M observations of GJ 876, we observed a flare in several lines formed in the active upper atmosphere of the star.  The light curves were extracted from the calibrated three-dimensional data by exploiting the time-tag capability of the COS microchannel plate detector~\citep{france10b}.  We extract a [$\lambda_{i}$,$y_{i}$,$t_{i}$] photon list (where $\lambda$ is the wavelength of the photon, $y$ is the cross-dispersion location, and $t$ is the photon arrival time) from each exposure $i$ and coadd these to create a master [$\lambda$,$y$,$t$] photon list. The total number of counts in a [$\Delta$$\lambda$,$\Delta$$y$] box is integrated over a timestep $\Delta$$t$.  The instrumental background level is integrated over the same wavelength interval as the emission lines, but offset below the active science region in the cross-dispersion direction.

Figure 4 shows the emission line count rates  
evaluated in $\Delta$$t$ = 40s timesteps.  The count rate in the emission lines at exposure times $T_{exp}$~$>$~4800s is decaying from the flare maximum.  The peak of the flare most likely occurred while the target was occulted by the Earth (the break in the observations between $T_{exp}$~$\approx$~1300~--~4800 s), which prevents us from measuring the actual peak or the full decay time of the flare.  However, the $observed$ flare maximum is approximately 10$\times$~the quiescent level, with an exponential decay time scale $\sim$~400s, implying a peak flare/quiescent FUV flux ratio of ~$\geq$~10.  Our data do not have the temporal coverage to determine the FUV flare frequency on GJ 876, but one could expect flares of this magnitude on timescales of hours to days~\citep{kowalski09}.  
We observe a weak FUV continuum associated with the flare event.  Integrating the total number of counts over an emission line-free 15~\AA\ region centered near~$\lambda$~$\sim$~1380~\AA, we find an orbit-averaged continuum/background ratio of 1.21~$\pm$~0.08.

\section{Discussion}

\subsection{The Importance of Ly$\alpha$ for Atmospheric Chemistry}  

Figure 1 and Table 1 show that the FUV/NUV ratio is much larger for M-dwarfs than for solar-type stars, and that the intrinsic 
Ly$\alpha$ flux dominates the HZ UV spectrum of GJ 876.   Comparing the intrinsic Ly$\alpha$ flux to the rest of the FUV (1150~--~1210 + 1220~--~1790\AA),  we see that the Ly$\alpha$/(FUV) ratio is $\sim$~8$\times$ the solar value.  The reconstructed Ly$\alpha$ emission line has roughly the same integrated flux as the rest of the UV spectrum (1150~--~3140~\AA) combined.  This is striking when compared to the relative contribution of Ly$\alpha$ to the total UV luminosity (Ly$\alpha$/(FUV+NUV) ) from the Sun.  GJ 876 displays a Ly$\alpha$/(FUV+NUV) ratio $\sim$~2500$\times$ larger than the solar value.

\begin{figure}[b]
\begin{center}
\epsfig{figure=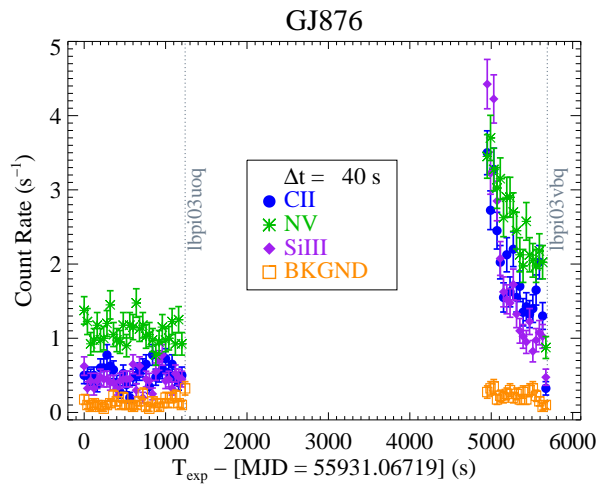,width=2.6in,angle=90}
\vspace{-0.2in}
\caption{
\label{cosovly} Direct observation of an FUV flare on the so-called ``inactive'' M-dwarf GJ 876 with $HST$-COS.  The count rates in various chromospheric and transition region emission lines~--~C II $\lambda$1335\AA [{\it blue circles}], N V $\lambda$1240\AA [{\it green stars}], Si III $\lambda$1206\AA [{\it purple diamonds}]~--~are displayed versus the time of observation.   The peak observed flare emission is $\approx$~10$\times$ the quiescent level, although the true flare peak (which apparently occurred during Earth occultation) could have been much larger.  
The detector background level is shown as orange squares.  
 }
\end{center}
\end{figure}

Because most of the UV luminosity from M-dwarfs is emitted in emission lines, molecular species in the atmospheres of orbiting planets will be ``selectively pumped'';  only species that spectrally coincide with stellar UV emission lines will be subject to large energy input from the host star.  
In order to estimate of the impact of Ly$\alpha$ irradiation, we computed the total flux absorbed by several important atmospheric constituents, $F_{\lambda, abs}$($X$) = $F_{FUV}$(GJ 876)~$\times$~(1~--~$e^{-\tau_{\lambda}(X)}$), where ($\tau_{\lambda}$($X$)) is calculated from published photoabsorption cross-sections for $X$~=~ CO$_{2}$~\citep{yoshino96}, H$_{2}$O~\citep{mota05}, or CH$_{4}$~\citep{lee01}. 
We find that Ly$\alpha$ photons account for $>$ 40\% of the energy absorbed by CO$_{2}$ and $>$ 85\% of the energy absorbed by H$_{2}$O and CH$_{4}$.  Other chromospheric and transition region emission lines contribute only a few percent to the energy absorbed by H$_{2}$O and CH$_{4}$.    \ion{C}{2}, \ion{Si}{4}, and \ion{C}{4} each contribute $\sim$~10\% to the FUV stellar flux absorbed by CO$_{2}$.  
While we are not presenting a rigorous treatment of the radiative transfer of Ly$\alpha$ photons incident on a planetary atmosphere (this will be done in future modeling work), this rough calculation highlights the importance of Ly$\alpha$ emission to the total energy budget available to drive atmospheric chemistry on these worlds.  

Strong FUV flux should be included in models of terrestrial planets in the HZ of ``weakly active'' and perhaps ``inactive'' M-dwarf host stars. Stellar FUV flux will drive CO$_{2}$ photolysis, leading to an enhancement of atmospheric O$_{2}$ (and O$_{3}$), although the exact enhancement factor must be determined.  \citet{segura07} showed that O$_{2}$ and O$_{3}$ column densities are enhanced by a factor of $\sim$~3 in CO$_{2}$-rich atmospheres when the FUV-bright young solar analog EK Dra is used as the illuminating radiation field.    In a terrestrial atmosphere illuminated by an older solar-type star, the much stronger NUV flux would serve to balance the possible enhancement of O$_{3}$ (the O$_{3}$ photolysis threshold is at $\lambda$~$\sim$~2500~\AA).  However the weak NUV field of GJ 876 should not destroy significant amounts of O$_{3}$.   We therefore suggest that the relatively strong FUV flux may increase the amount of atmospheric O$_{3}$ present relative to what is predicted by models of terrestrial HZ planets orbiting ``inactive'' M-dwarfs~\citep{segura05}.   
This highlights the need for measured UV spectra of the host star when modeling the atmospheric properties of an exoplanet.  The paucity of UV observations of low-mass exoplanetary host stars may bias our interpretation of potential biomarkers when direct spectroscopy of extrasolar terrestrial planets becomes feasible, possibly in the coming decade with the {\it James Webb Space Telescope}~\citep{belu11}. 

\subsection{Fluorescent Emission from Hot H$_{2}$}  

The COS spectra of GJ 876 show surprisingly strong emission from fluorescent H$_{2}$, excited by Ly$\alpha$ photons.  
Jupiter's dayglow and auroral spectra are dominated by collisionally-excited and Ly$\beta$-pumped H$_{2}$~\citep{feldman93,gustin02}.  However, emission from hot H$_{2}$ pumped by Ly$\alpha$ photons was observed at the Shoemaker-Levy 9 impact site~\citep{wolven97}.  If there were a heat source capable of producing high molecular temperatures ($T$(H$_{2}$)~$>$~2000 K) in the atmosphere of GJ 876b or GJ 876c, then the observed fluorescence could have an exoplanetary, rather than a stellar origin.  Similar emission was searched for from the well-studied transiting planet HD 209458b, but not conclusively detected~\citep{france10a}.  

We cannot rule out the possibility that some of the H$_{2}$ emission is associated with GJ 876b or GJ 876c, but a calculation of the available energy budget argues for a stellar origin for this fluorescence.  From the reconstructed Ly$\alpha$ profile (\S3.2), we can calculate the Ly$\alpha$ surface flux, $F_{surf}$(Ly$\alpha$)~=~$F_{obs}$(Ly$\alpha$) $\times$~($d$/$R_{*}$)$^{2}$, where $R_{*}$ is the stellar radius $R_{*}$~$\approx$~0.3 $R_{\odot}$.   The Ly$\alpha$ flux absorbed by H$_{2}$ is then $F_{abs}$(H$_{2}$)~=~$F_{surf}$(Ly$\alpha$)~$\times$~(1~--~$e^{-\tau_{\lambda}(H_{2})}$), where $\tau_{\lambda}$(H$_{2}$) is the optical depth of H$_{2}$, calculated for $N$(H$_{2}$)~=~10$^{17}$ cm$^{-2}$ (the absorbed flux only increases by a factor of $\sim$~2.5 when $N$(H$_{2}$)~=~10$^{19}$ cm$^{-2}$) and $T$(H$_{2}$)~=~2500 K.   If we assume that the H$_{2}$ resides in an optically thin circular cloud, the emitted and absorbed luminosities will be equal, $L_{abs}$(H$_{2}$) = $\pi$$R_{H2}^2$$F_{abs}$(H$_{2}$)~=~$L_{emis}$(H$_{2}$).  Focusing on the [$v^{'}$,$J^{'}$] = [1,4] progression, $L_{emis}$(H$_{2}$)~=~4$\pi$$d^{2}$~$\times$~($F_{(1-7)P(5)}$/$B_{17}$), where $F_{(1-7)P(5)}$ is the flux in the H$_{2}$ (1~--~7)P(5) $\lambda$1504.76 emission line ($F_{(1-7)P(5)}$~=~(2.2 $\pm$ 0.4)~$\times$~10$^{-16}$ erg cm$^{-2}$ s$^{-1}$) and $B_{17}$ is the $v^{'}$~$\rightarrow$~$v^{''}$ = 1~$\rightarrow$~7 branching ratio ($B_{17}$~=~0.115).  Equating $L_{emis}$(H$_{2}$) and $L_{abs}$(H$_{2}$), we find $R_{H2}$~=~0.32~$\pm$~0.13~$R_{\odot}$, consistent with the H$_{2}$ being located in the stellar atmosphere.  A similar calculation at the semi-major axis of GJ 876b would require $R_{H2}$~$>$~400$R_{Jup}$, much larger than observational or theoretical estimates of the size of hot Jupiter envelopes~\citep{madjar08, murray09}.   We conclude that the observed fluorescence is produced by hot photospheric molecules illuminated by the strong chromospheric Ly$\alpha$ radiation field.   \\ \\ 

\acknowledgments
K.F. thanks Sarah LeVine for graphic support of Project MUSCLES.  We acknowledge enjoyable discussions with John Stocke, Hao Yang, and Brian Wolven.  
These data were obtained as part of the $HST$ Guest Observing program \#12464.  This work was supported by NASA grants NNX08AC146 and NAS5-98043 to the University of Colorado at Boulder.  




\end{document}